\documentclass[a4paper]{article}

\usepackage{INTERSPEECH2020}
\usepackage{multirow}
\usepackage{algorithm}
\usepackage{algorithmic}

\title{U2++: Unified Two-pass Bidirectional End-to-end Model for Speech Recognition}
\name{Di Wu$^{1,2}$, Binbin Zhang$^{1,2}$, Chao Yang$^{1,2}$, Zhendong Peng$^{1,2}$, Wenjing Xia$^1$, Xiaoyu Chen$^1$,  Xin Lei$^1$}

\address{
  $^1$Mobvoi Inc., Beijing, China \\
  $^2$WeNet Open Source Community}
\email{di.wu@mobvoi.com 
       binbinzhang@mobvoi.com 
       chaoyang@mobvoi.com 
       zhendong.peng@mobvoi.com
       wenjing.xia@mobvoi.com
       xiaoyu.chen@mobvoi.com
       mikelei@mobvoi.com}

\begin{document}

\maketitle
\begin{abstract}

The unified streaming and non-streaming two-pass (U2) end-to-end model for speech recognition has shown great performance in terms of streaming capability, accuracy, real-time factor (RTF), and latency.
In this paper, we present U2++, an enhanced version of U2 to further improve the accuracy.
The core idea of U2++ is to use the forward and the backward information of the labeling sequences at the same time at training to learn richer information, and combine the forward and backward prediction at decoding to give more accurate recognition results.
We also proposed a new data augmentation method called SpecSub to help the U2++ model to be more accurate and robust.
Our experiments show that, compared with U2, U2++ shows faster convergence at training, better robustness to the decoding method, as well as consistent 5\% - 10\% word error rate reduction gain over U2.
On the experiment of AISHELL-1, we achieve a 4.63\% character error rate (CER) with a non-streaming setup and 5.05\% with a streaming setup with 320ms latency by U2++.
To the best of our knowledge,  5.05\% is the best-published streaming result on the AISHELL-1 test set.

\end{abstract}

\section{Introduction}

End-to-end (E2E) models such as CTC\cite{graves2006connectionist,amodei2016deep}, recurrent neural network transducer (RNN-T)\cite{graves2012sequence,graves2013speech}, and attention-based encoder-decoder (AED)\cite{chorowski2014end,chan2015listen,chorowski2015attention} gained more and more attention over the last few years.
Compared with the conventional hybrid ASR framework, E2E models not only extremely simplified training and decoding procedure but also show superior performance in the standard of word error rate (WER).
So applying the E2E models into real-world productions becomes necessary. However, deploying an E2E system is not trivial and there are a lot of practical problems to be solved.

The first is the streaming problem, some state-of-the-art models such as vanilla AED could not run in a streaming way, which limits the application scenarios. The second is that streaming and non-streaming systems are usually developed, trained, and deployed separately, which is resource-consuming.
The third is the production problem, a lot of engineer efforts are required to promote e2e models to the production level.

Our previously proposed U2\cite{zhang2020unified}, a unified streaming and non-streaming end-to-end model, and WeNet\cite{yao2021wenet}, a production first and production-ready E2E toolkit solve the above problems in a simple and graceful way.

In this paper, we present U2++, an enhanced version of U2 to further improve the performance.
Compared with U2, where there is only one left-to-right (forward) decoder, we add another right-to-left (backward) decoder. It means we have bidirectional decoders in U2++, which uses both the forward and the backward information of the labeling sequence.
In the training, the model is trained jointly by CTC loss, left-to-right AED loss, right-to-left AED loss, so it learns richer information than U2.
In the decoding, both the left-to-right decoder and the right-to-left decoder are used for re-scoring the n-best hypotheses from the CTC decoder, so we can make a more accurate prediction.
Besides, inspired by SpecSwap\cite{song2020specswap} and SpecAugment\cite{park2019specaugment}, we also proposed a novel data augmentation method called SpecSub, which randomly and dynamically replace the current feature block with the feature block before.
Our experiments show that U2++ is more stable, accurate, and robust than U2.
The CTC and bidirectional AED joint training benefits both the CTC result and the final re-scoring result.
On average, we get a 5\%-8\% relative word error rate reduction by U2++.
On the AISHELL-1 dataset\cite{bu2017aishell}, we get a 5.05\% CER on the test set with a streaming setting. As far as we know, it is the best streaming result on AISHELL-1.

\section{Related Works}
Hybrid CTC/attention end-to-end ASR\cite{kim2017joint} adopts both CTC and attention decoder loss during training, which results in faster convergence and also improves the robustness of the AED model.
The encoder module can use any neural network which can model contexts, such as CNN, LSTM, conventional Transformer encoder, and recently proposed Conformer encoder which gets SOTA results on many public data sets.
The decoder module of the hybrid CTC/attention model usually uses a conventional transformer decoder, in which left-to-right self-attention and across attention are used.

There are many decoding strategies of the hybrid CTC/attention model during the decoding process.
EspNet\cite{kim2017joint} combines CTC search and auto-regressive beam search decoding of the attention decoder.
U2 adopts a two-pass decoding strategy, the first pass is a CTC prefix beam search which gives the n-best candidate hypothesizes by CTC decoder, the second pass is a re-scoring process that re-scores the n-best candidates and the candidate with the best score is chosen to be the final result.

There are some works on bidirectional decoder on conventional speech transformer to improve the accuracy. \cite{chen2020transformer} adds a right-to-left decoder to the conventional speech transformer model, and both the left-to-right decoder and the right-to-left decoder use auto-regressive beam search during the decoding. \cite{higuchi2020mask} use a BERT style decoder instead of the conventional speech transformer so that it models both the left and the right context, but the decoding process is complicated.
Pika\footnote{https://github.com/tencent-ailab/pika}, a RNN-T based E2E toolkit, supports bidirectional LAS rescoring on RNN-T model, however, experimental results, experimental analysis and recipes are not given.

\section{U2++}

\subsection{Model architecture}
\begin{figure}[h]
  \centering
  \includegraphics[width=\linewidth]{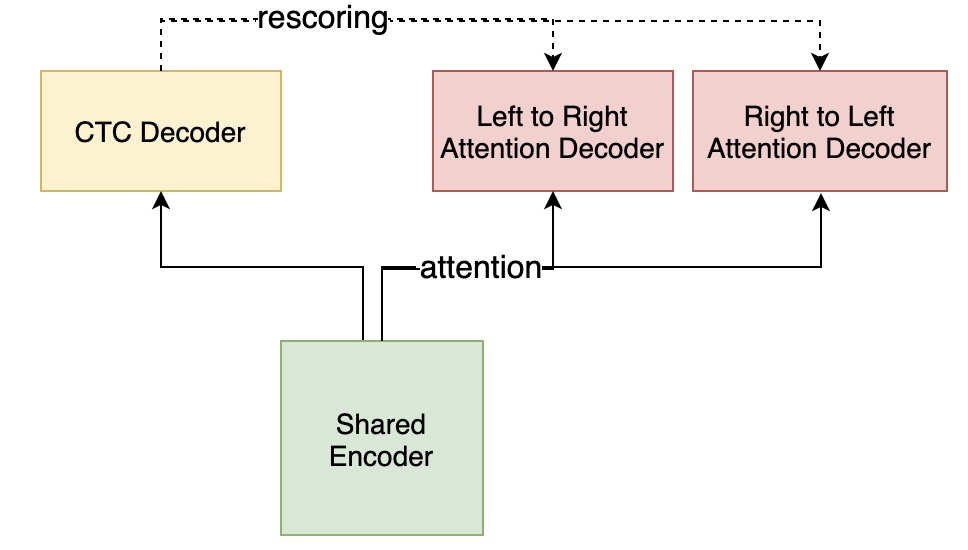}
  \caption{Two pass CTC and AED joint architecture}
  \label{fig:ctc_attention_joint}
  \vspace{-1.0em}
\end{figure}

The proposed model architecture is shown in Figure\ref{fig:ctc_attention_joint}. It contains four parts, a \textit{Shared Encoder} that models the context of the acoustic features, a \textit{CTC Decoder} that models the alignment of the frames and tokens, a \textit{Left-to-Right Attention Decoder (L2R)} that models the left tokens dependency, and a \textit{Right-to-Left Attention Decoder (R2L)} model the right tokens dependency. The \textit{Shared Encoder} consists of multiple Transformer\cite{vaswani2017attention} or Conformer\cite{gulati2020conformer} encoder layers. The \textit{CTC Decoder} consists of a linear layer and a log softmax layer. The CTC loss function is applied over the softmax output in training.

The \textit{Left-to-Right Attention Decoder} and the \textit{Right-to-Left Attention Decoder} use the same model structure as the conventional transformer decoder.   
We still make the \textit{Shared Encoder} only see limited right context which is the same as U2, then the CTC decoder could run in a streaming mode in the first pass. In the second pass, we still use the re-scoring method as used in U2, but since we have two decoders in U2++, the re-scoring process will be a little different from U2. The detailed training and decoding processes are described in the following.
\subsection{Training}
\subsubsection{Combined Loss}

The training loss is the combined CTC and AED loss as listed in the equation\ref{eq:combined_loss}, and the AED loss consists of an L2R AED loss and an R2L AED loss as listed in the equation \ref{eq:AED_loss}. Both of the two attention decoders are trained in the teacher forcing mode. In both equation \ref{eq:AED_loss} and equation \ref{eq:combined_loss}, $\mathbf{x}$ are the acoustic features, $\mathbf{y}$ are the corresponding annotations, $\mathbf{L}_{\text{CTC}}\left(\mathbf{x}, \mathbf{y}\right)$ and 
$\mathbf{L}_{\text{AED}}\left(\mathbf{x}, \mathbf{y}\right)$ are the CTC and AED loss respectively.
$\lambda$ is a hyperparameter that balances the importance of the CTC and AED loss. $\alpha$ is another hyperparameter that balances the R2L AED loss and the L2R AED loss. 

\begin{figure}[h]
  \centering
  \includegraphics[width=\linewidth]{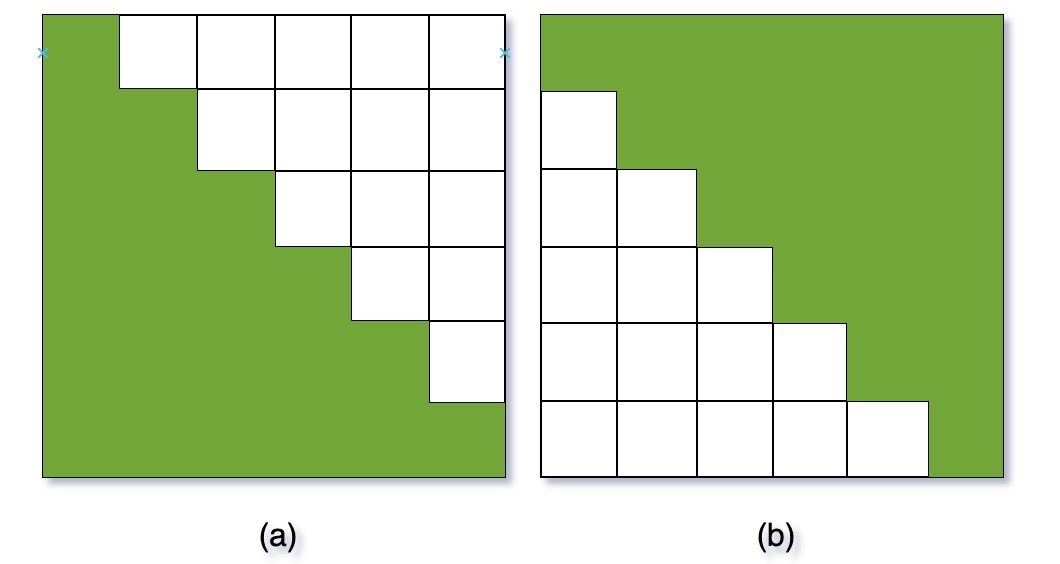}
  \caption{left attention mask and right attention mask}
  \label{fig:mask}
  \vspace{-2.0em}
\end{figure}

\begin{equation}
  \label{eq:combined_loss}
    \mathbf{L}_{\text {combined}}\left(\mathbf{x}, \mathbf{y}\right)=\lambda \mathbf{L}_{\text{CTC}}\left(\mathbf{x}, \mathbf{y}\right)+(1-\lambda) (\mathbf{L}_{\text{AED }}\left(\mathbf{x}, \mathbf{y}\right))
    \vspace{-2.0em}
\end{equation}

\begin{equation}
  \label{eq:AED_loss}
    \mathbf{L}_{\text {AED}}\left(\mathbf{x}, \mathbf{y}\right)=(1 - \alpha) \mathbf{L}_{\text{AED-L2R}}\left(\mathbf{x}, \mathbf{y}\right) + \alpha (\mathbf{L}_{\text{AED-R2L}}\left(\mathbf{x}, \mathbf{y}\right) )
\end{equation}

\subsubsection{Bi-directional Attention Decoder}

In \cite{chen2020transformer}, the L2R and R2L decoders share the same weights. In the decoding process, these two decoders run separately and the sequence with the best score is chosen as the final result. Although This strategy could give more candidates, it is hard to combine the two direction scores due to the autoregressive mode. 

Our models have mainly two differences from \cite{chen2020transformer} . First, we used different weights for the two decoders. Because the reversed version language should not be considered as the reasonable original language. Second, in the U2++ framework, the attention decoders are used for re-scoring instead of autoregressive decoding, so the two decoder scores of the candidates could be combined.

The implementations are described in the following. The mask in Figure 2(a) is applied to the self-attention in the L2R decoder to make each word only see its left words. Similarly, The mask in Figure 2(b) is applied in the R2L decoder to make each word only see its right words. Another implementation is to reverse the input sentence and use Figure 2(a) mask. The two implementations are only different in the position encoding information and give almost the same performance in experiments. So in this paper, we only report the result of the first implementation.

\subsubsection{Spectral Augmentation}

We propose a new spectral augmentation method called SpecSub to make the model more robust. We randomly substitute a frame chunk with the previous chunk. This simple augmentation shows significant performance improvements in the experiments.

\begin{algorithm}
	\caption{SpecSub Augmentation}
	\label{alg1}
	\begin{algorithmic}[1]
		\STATE Initialization: $ T_{max}, T_{min}, N_{max}, n=1$
		\STATE sample $ N \sim  Uni(0, N_{max}) $
		\FOR{n from 1 to N}
            \STATE sample $ \Delta_{t} \sim  Uni(T_{min}, T_{max}) $
    		\STATE sample $ t \sim  Uni(0, T - \Delta_{t}) $
    		\STATE sample $ t' \sim  Uni(0, t) $
    		\STATE $f(t, t +\Delta_{t} ) \leftarrow f(t', t' +\Delta_{t} )$
		\ENDFOR
	\end{algorithmic}  
\end{algorithm}

\subsection{Decoding}

We use the similar two-pass re-scoring decoding strategy with U2, while the new R2L decoder score is involved in re-scoring in U2++. 

In the first pass stage, the frame-sync CTC decoder outputs the candidates using prefix beam search. In the second pass stage, the two attention decoder scores of these n-best candidates are calculated. The final score is calculated via equation \ref{eq:combined_scores}. The candidate with the best score is chosen as the final result.
\begin{equation}
  \label{eq:combined_scores}
    \mathbf{S}_{\text {final}} =\lambda\times \mathbf{S}_{\text {CTC}} + (1 - \alpha) \times \mathbf{S}_{\text {L2R}} + \alpha\times \mathbf{S}_{\text {R2L} }
\end{equation}

\section{Experiments}

To evaluate the proposed U2++ model, we carry out our experiments on the open-source Chinese Mandarin speech corpus AISHELL-1\cite{bu2017aishell} and the open-source Chinese Mandarin speech corpus AISHELL-2. AISHELL-1 contains a 150-hour training set, a 20-hour development set, and a 10-hour test set. The test set contains 7176 utterances in total. AISHELL-2 contains a 1000-hour training set, and we use $dev\_ios$ as a development set and $test\_ios$ that contains 5000 utterances in total as the test set for validation and evaluation. We use WeNet\footnote{https://github.com/mobvoi/wenet} - an end-to-end speech recognition toolkit for all experiments. 

The 80-dimensional log Mel-filter banks (FBANK) computed on-the-fly by torchaudio with a 25ms window and a 10ms shift are used as the features.
SpecAugment~\cite{park2019specaugment} is applied with two frequency masks with maximum frequency mask ($F = 10$), and two times masks with maximum time mask ($T=50$) to alleviate the over-fitting. The $T_{max}$, $T_{min}$, $N_{max}$ of the proposed SpecSub is set to 30, 0 and 3 respectively.

We use two state-of-the-art ASR networks Conformer and Transformer as our shared encoder to evaluate U2++.
We replace the batch-norm in the convolution module of Conformer with layer-norm, and we use the causal convolution in the convolution module.
The relative shift in the original Conformer is removed for streaming.
Two convolution sub-sampling layers with kernel size 3*3 and stride 2 are used in the front of the encoder.
The kernel size of the convolution is 8 in Conformer.
We use 12 transformer/conformer layers for the encoder.
3 transformer layers are used for both the L2R decoder and the R2L decoder, and the number of attention heads, attention dimension, and feed-forward dimension are set 4, 256, and 2048 respectively. So the parameters in U2++ are almost the same as U2.

Adam optimizer is used and the learning rate is warmed up with 25,000 steps. The $\alpha$ and $\lambda$ are both set to 0.3 in training. In decoding, $\alpha$ and $\lambda$ are set according to the developing set performance. In AISHELL-1 task, they are 0.3 and 0.5. In AISHELL-2 task, they are 0.1 and 0.5. The final model averages the top 30  models with the best validation loss on the development set. The beam size of the CTC prefix search is ten.

\begin{table}[h]
\footnotesize
    \vspace{-5pt}
    \centering
    \setlength{\abovecaptionskip}{0.1cm}
    \caption{Test set CER (\%) of U2/U2++ trained on AISHELL-1 Mandarin speech data.}
    \label{tab:aishell u2++}
    \begin{tabular}{c c c c c}
    \toprule
\multirow{2}{*}{training method}  & \multirow{2}{*}{decoding mode} & \multicolumn{2}{l}{decoding chunk size} \\ \cline{3-4} 
                               & & full   & 16          \\ \hline
    \multirow{2}{*}{U2 Transformer}    & ctc prefix beam search         & 6.28   & 6.98    \\ 
    & attention rescoring            & 5.52   & 6.05    \\ 
    \hline
    \multirow{2}{*}{U2++ Transformer}    & ctc prefix beam search         & 6.05   & 6.92    \\ 
    & attention rescoring            & 5.09   & 5.66   \\ 
    \hline
    \multirow{2}{*}{U2 Conformer}     & ctc prefix beam search         & 5.57   & 6.30   \\ 
    & attention rescoring            & 4.97   & 5.45   \\ 
    \hline
    \multirow{2}{*}{ U2++ Conformer }   & ctc prefix beam search         & 5.19   & 5.81 \\ 
    & attention rescoring            & 4.63   & 5.05  \\ 

    \hline
\end{tabular}%
    \vspace{-15pt}
\end{table}

\subsection{AISEHLL-1 Task}

\begin{table}[h]
    \vspace{-5pt}
    \setlength{\abovecaptionskip}{0.1cm}
    \centering
    \caption{Comparison to other streaming solutions on AISHELL-1 task}
    \label{tab:solution_comparison}
    \begin{tabular}{lllll}
    \toprule
    model            & latency(ms) & lm & CER  \\ \midrule
    Sync-Transformer\cite{tian2020synchronous}     & 400  & n & 8.91 \\
    SCAMA\cite{zhang2020streaming}                & 600   & n & 7.39 \\
    MMA\cite{inaguma2020enhancing}                & 640  & n & 6.60  \\
    U2 \cite{zhang2020unified}             & \textbf{320}+$\Delta$  & n & \textbf{5.33} \\
    WNARS \cite{wang2021wnars}           & 640          & y & \textbf{5.15} \\
    U2++              & \textbf{320}+$\Delta$  & n & \textbf{5.05} \\
    \bottomrule
    \end{tabular}
        \vspace{-5pt}
\end{table}

We use the same dynamic chunk training as U2, which enables the model to work well in different chunk sizes (full and 16). The full chunk is for a non-streaming manner and chunk=16 is for streaming. the model latency of chunk=16 mode is ranged from 0 to 640ms, so the average latency is 320ms. Speed perturbation with 0.9, 1.0, and 1.1 is applied on the fly on the whole data.

\begin{figure}[h]
  \centering
  \includegraphics[width=\linewidth]{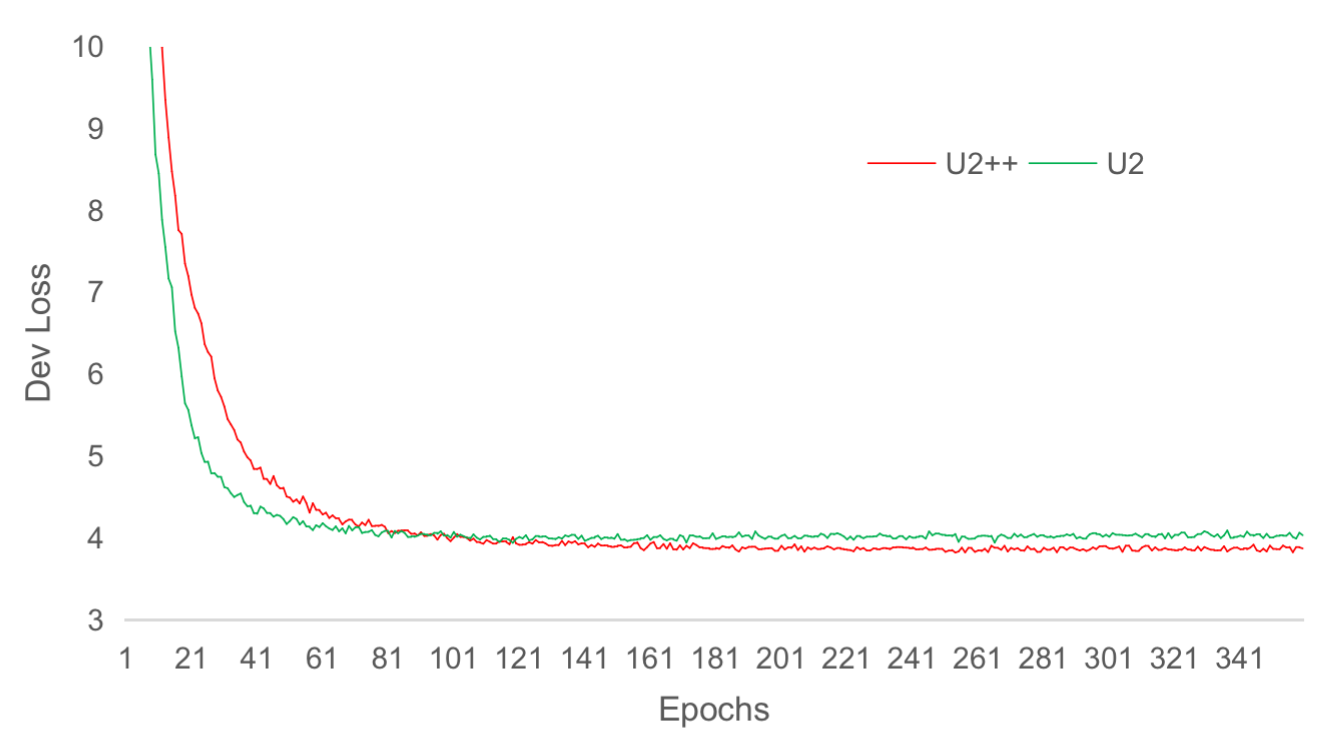}
  \caption{The loss comparison of U2++ and U2}
  \label{fig:loss}
  \vspace{-1.0em}
\end{figure}

The model performance comparisons of U2 and U2++ are shown in Table \ref{tab:aishell u2++}, U2++ is better than U2 on both Transformer and Conformer model in both streaming and non-streaming mode. The U2++ and U2 Conformer model losses are shown in Figure \ref{fig:loss}. The loss of U2++ is higher than U2 in the early stage, but it converges to lower after enough epochs.

Table \ref{tab:solution_comparison} shows the comparisons with the other published systems on this task. 'y' in 'lm' column means a separated language model is used and 'n' means not.
The averaged latency of the U2 and U2++ model is about 320ms and $\Delta$ usually is about 100ms.
U2++ also achieves the best CER results in streaming mode on the AISHELL-1 data set and even better than WNARS which uses a separated LM during the decoding process.

\subsection{AISEHLL-2 Task}
We extend our experiments on the larger AISHELL-2 dataset. In AISHELL-2, speed perturbations are not applied. The performance of U2++ is listed in Table \ref{tab:15000 hour}. On this dataset, U2++ still does better than U2 both on the Transformer and the Conformer model.
\begin{table}[h]
\footnotesize
    \vspace{-5pt}
    \centering
    \setlength{\abovecaptionskip}{0.1cm}
    \caption{AISHELL-2 Mandarin speech data trained U2 and U2++, tested on test set in CER (\%). }
    \label{tab:15000 hour}
    \begin{tabular}{c c c c c}
    \toprule
\multirow{2}{*}{training method}  & \multirow{2}{*}{decoding mode} & \multicolumn{2}{l}{decoding chunk size} \\ \cline{3-4} 
                               & & full   & 16          \\ \hline
    \multirow{2}{*}{U2 Transformer}    & ctc prefix beam search         & 7.84   & 8.68    \\ 
    & attention rescoring            & 6.71   & 7.31    \\ \hline
    \multirow{2}{*}{U2++ Transformer}    & ctc prefix beam search         & 7.45   & 8.21   \\ 
    & attention rescoring            & 6.09   & 6.70   \\ \hline
    \multirow{2}{*}{U2 Conformer}     & ctc prefix beam search         & 7.02   & 7.76   \\ 
    & attention rescoring            & 6.08   & 6.46   \\ \hline
    \multirow{2}{*}{ U2++ Conformer }   & ctc prefix beam search         & 6.70   & 7.39 \\ 
    & attention rescoring            & 5.39   & 5.78  \\
    \hline
\end{tabular}%
    \vspace{-15pt}
\end{table}
\begin{table}[h]
    \vspace{-5pt}
    \setlength{\abovecaptionskip}{0.1cm}
    \centering
    \caption{Ablation study of U2++, we remove the additional modules compared to the U2 model:(1) only removing R2L decoder; (2) only removing SpecSub ;(3)  removing R2L decoder and SpecSub.  }
    \label{tab:proposed_comparison}
    \begin{tabular}{lll}
    \toprule
    model &   full chunk & chunk 16\\ \midrule
    u2++ baseline  & 4.63 & 5.05\\
    \quad- R2L decoder  & 4.78 & 5.25\\
    \quad- SpecSub         & 4.78 & 5.22\\
    \quad\quad- R2L decoder (U2) & 4.97 & 5.45 \\
    \bottomrule
    \end{tabular}
    \vspace{-5pt}
\end{table}
\subsection{Ablation Study}

\subsubsection{Proposed Improvement}

We evaluate the effectiveness of the proposed SpecSub and the added R2L decoder loss. The results are shown in the Table \ref{tab:proposed_comparison}.Both of them bring a significant improvement on the U2 baseline and the results are better when they are combined. 
\subsubsection{Individual Decoder Using in Decoding Process}
\begin{table}[t]
    \vspace{-1pt}
    \setlength{\abovecaptionskip}{0.1cm}
    \centering
    \caption{Ablation study of the individual L2R decoder and R2L decoder using for re-scoring (RS) and auto-regressive beam search decoding method (AR) on AISHELL-1 test set. For U2: (1) reducing three decoder layers directly during training. For U2++: (1) only using R2L decoder during decoding; (2) only using L2R decoder during decoding. The results with same * or † denotes that it can be fair to directly compare.}
    \label{tab:two_decoder_comparison}
    \begin{tabular}{llll}
    \toprule
    model & method & full chunk & chunk 16\\ \midrule
    \multirow{2}{*}{U2 6layer
              *} &  AR & 5.19 & 5.47 \\
             &  RS      & 4.97 & 5.45\\
    \multirow{2}{*}{U2 3layers † }
         &  AR   & 5.27 & 5.55\\
         &  RS   & 4.99 & 5.49\\ \hline 
    \multirow{2}{*}{U2++ baseline * } 
    &  AR  & 4.83 & 5.12\\
    &  RS  & 4.63 & 5.05\\
   \multirow{2}{*}{only R2L decoder †}          &  AR  & 5.07 & 5.35 \\  
   &  RS  & 4.81 & 5.19 \\
    \multirow{2}{*}{only L2R decoder †}   
    &  AR  & 5.01 & 5.30 \\
    &  RS   & 4.76 & 5.15  \\
    \bottomrule
    \end{tabular}
    \vspace{-10pt}
\end{table} 

We also evaluate the effectiveness of the two decoders in U2++ by using three different re-scoring methods, which are only the L2R score, only the R2L decoder score, and the combined score. The layer number of the U2 decoder is 6. To keep the parameter the same as U2++,  we also train a U2 model with 3 decoder layers. In addition, we compare our re-scoring strategy with the auto-regressive strategy used in \cite{chen2020transformer} 
 in all settings. The details are reported in the Table \ref{tab:two_decoder_comparison}. 

The re-scoring decoding method still shows better results than the auto-regressive beam search in the U2++ model. No matter which decoding method is used, U2++ is better than U2 with three or six decoder layers. Even with only single decoders, the U2++ performs better than U2 with six decoder layers. When combining the scores of two direction decoders, U2++ achieves 7\% relative CER  reduction with U2.

The U2++ also shows the ability to make a trade-off between RTF/latency and CER. When lower RTF/latency is the top priority, users could use a single L2R decoder in exchange for better RTF.

\begin{table}[h]
    \vspace{-1pt}
    \setlength{\abovecaptionskip}{0.1cm}
    \centering
    \caption{Study on $\alpha$ value during the training process on AISHELL-1 test set, All the results are conduct by Conformer U2++ model.}
    \label{tab:reverse_comparison}
    \begin{tabular}{lll}
    \toprule
    weight $\alpha$ &   full chunk & chunk 16\\ \midrule
    0.1   & 4.74 & 5.09\\
    0.3   & 4.63 & 5.05 \\
    0.5   & 4.67 & 5.13  \\
    \bottomrule
    \end{tabular}
    \vspace{-1.0em}
\end{table}

\begin{table}[!h]
    \vspace{-1pt}
    \setlength{\abovecaptionskip}{0.1cm}
    \centering
    \caption{Study on the number of R2L and L2R decoder layers on AISHELL-1 test set.}
    \label{tab:decoder_layer_comparison}
    \begin{tabular}{lll}
    \toprule
    decoder layers &   full chunk & chunk 16\\ \midrule
    L2R=5, R2L=1   & 4.74 & 5.09\\
    L2R=4, R2L=2   & 4.75 & 5.13 \\
    L2R=3, R2L=3   & 4.63 & 5.05  \\
    \bottomrule
    \end{tabular}
    \vspace{-1.5em}
\end{table}

\subsubsection{Score Weights} 
The weight $\alpha$ is studied with different values during the training process include 0.1, 0.3, 0.5. We search the $\alpha$ and $\lambda$ during the decoding process.
The influence of $\alpha$ during the training process is detailed in Table \ref{tab:reverse_comparison}.
While $\alpha$ is set 0.3 during training, the decoding result is better than the rest.

\subsubsection{Allocation of Decoder Layer}
In a fixed total parameters numbers setting,  the parameters allocated to L2R and R2L decoders may influence the model performance. Our experiments on a total of six layers show that allocating the number of the parameters equally on the two decoders performs best. The details are reported in Table \ref{tab:decoder_layer_comparison}. 

\section{Conclusions}
 In this paper, we analyze the shortcoming of U2, which is the lack of modeling the right words in the attention decoder. So we propose the R2L decoder to involve the right information. We also proposed the SpecSub data augmentation method to make the model trained by dynamic chunk training more robust. Future works include exploring the decoder model consistent in the training and decoding process, such as a non-aggressive attention-based decoder. And more experiments are needed on the larger industrial datasets.

\newpage

\bibliographystyle{IEEEtran}
\bibliography{main}

\end{document}